\documentclass[a4paper,12pt]{article}
\usepackage{epsfig}
\usepackage{graphics}
\usepackage{amsmath}
\usepackage{amsfonts}

\newcommand{\n}{\mathbf{n}}
\usepackage{amssymb}
\usepackage{amstext}
\usepackage{morefloats}
\begin{document}

\title{Entropy as a function of Geometric Phase}%
\author{Julian Hartley and Vlatko Vedral}%
\maketitle
\begin{abstract}
We give a closed-form solution of von Neumann entropy as a
function of geometric phase modulated by visibility and average
distinguishability in Hilbert spaces of two and three dimensions.
We show that the same type of dependence also exists in higher
dimensions. We also outline a method for measuring both the
entropy and the phase experimentally using a simple Mach-Zehnder
type interferometer which explains physically why the two concepts
are related.
\end{abstract}

\section{Introduction}
The von Neumann entropy \cite{vonneumann} is a measure of
mixedness in a physical state described by a density matrix. The
general rule is that the more orthogonal the states comprising the
density matrix are, the higher the value of the corresponding
entropy. Looking at it from a different perspective, the entropy
signifies the lack of knowledge we have about the exact pure state
the system is in. For pure states, the knowledge is maximal and
the value of entropy is zero, while for a maximally mixed state
(the normalized identity matrix), the value of entropy is highest
as any of the pure states in the mixture is equally likely.
Therefore, this intuition would suggest that distinguishability
between states is the only parameter determining the value of
entropy. We also note that entropy is a static property of the
system (i.e. it is only a function of the density matrix
describing the state, rendering it completely insensitive to the
dynamical evolution).

Geometric phases, on the other hand, are obtained when a physical
system evolves through a (discrete or continuous) set of states.
We can say that this phase depends only on the geometric aspects
of this evolution (i.e. it is, for instance, independent of the
rate of evolution), but that it is still a dynamical property of
the system. In other words, it is generated by dynamics, although
the dynamics can be either a continuous Schr\"{o}dinger type
evolution or a discrete quantum measurement (of the most general
type). The geometric phase has a long and interesting history, and
we refer the interested reader to the collection of papers
compiled by Shapere and Wilczek \cite{wilczek}. No detailed
knowledge of this will be necessary however, as all the relevant
information will be given here.

Given that the entropy is a static property and geometric phase a
dynamical property of a quantum system, we would not at first
sight expect there to be any connections between the two. This
conclusion is however incorrect and in this paper, we will show
that entropy can in fact be written as a function of geometric
phase (and some other parameters in general).

Our work has been stimulated by Jozsa and Schlienz
\cite{jozsaschlienz} who pointed out that von Neumann entropy can
increase even when the ensemble of quantum states become less
distinguishable (i.e. more parallel). They noticed that this
behaviour does not occur in a two dimensional Hilbert space but
emerges in a three dimensional Hilbert space. Their conclusion is
that distinguishability is a global property (considering the
whole ensemble) which cannot be reduced to considering the
pairwise overlaps of the states.

In this paper, we attribute this transition to the presence of a
geometric phase by giving a closed-form solution of entropy as a
function of geometric phase. We will begin by defining all the
relevant variables. Then we will work through the two and three
dimensional cases. We also comment on the arbitrary dimensional
case. We will finally briefly discuss a method to experimentally
measure entropy and show that the same set up is also used for
measuring geometric phases. It is for this reason precisely that
the two concepts are related. Interestingly, in two dimensions,
the entropy is either a function of the phase or
distinguishability, but we do not need both at the same time (this
is because the phase and distinguishability can directly be
related to each other). For higher dimensions this relationship
becomes more complicated as we will show and throughout the paper
we discuss mathematical and physical reasons for this difference
between two and higher dimensional systems. We will conclude by
discussing the implications of our results with possible
generalizations.

\section{Setting the Scene}

As we have already said, entropy is a physical quantity that
quantifies the lack of information in a given ensemble. Suppose
the ensemble contains three quantum states $|\psi_1\rangle,
|\psi_2\rangle, |\psi_3\rangle$ with prior probabilities $p_1,
p_2, p_3$ respectively where $p_1+p_2+p_3=1$. We can construct the
density operator
$\rho=p_1|\psi_1\rangle\langle\psi_1|+p_2|\psi_2\rangle\langle\psi_2|+p_3|\psi_3\rangle\langle\psi_3|$
and the von Neumann entropy as $S_{vN}=-Tr(\rho\ln\rho)$ where the
Boltzmann constant $k_B=1$. Rather than state vectors, we shall be
predominantly working with coherence vectors which is a completely
analogous description. This is because we can generalize to higher
number of states than the dimension of the system. Any density
operator for two dimensions can be written as
$\rho=\frac{1}{2}(\mathbb{I}+\n\cdot\mathbf{\sigma})$, where
$\sigma$ are the Pauli matrices:
\begin{equation}\label{pauli}
    \sigma_1=\left(%
\begin{array}{cc}
  0 & 1 \\
  1 & 0 \\
\end{array}%
\right), \sigma_2=\left(%
\begin{array}{cc}
  0 & -i \\
  i & 0 \\
\end{array}%
\right), \sigma_3=\left(%
\begin{array}{cc}
  1 & 0 \\
  0 & -1 \\
\end{array}%
\right)
\end{equation}
$\n$ is a three component coherence vector and $\cdot$ denotes the
scalar product. In three dimensions any state can be written as
$\rho=\frac{1}{3}(\mathbb{I}+\sqrt{3}\n\cdot\mathbf{\lambda})$
where $\lambda$ are the Gell-Mann matrices:
\begin{align}\label{gellmann}
    &\lambda_1=\left(%
\begin{array}{ccc}
  0 & 1 & 0 \\
  1 & 0 & 0 \\
  0 & 0 & 0 \\
\end{array}%
\right),
&\lambda_2=\left(%
\begin{array}{ccc}
  0 & -i & 0 \\
  i & 0 & 0 \\
  0 & 0 & 0 \\
\end{array}%
\right), \quad \quad
&\lambda_3=\left(%
\begin{array}{ccc}
  1 & 0 & 0 \\
  0 & -1 & 0 \\
  0 & 0 & 0 \\
\end{array}%
\right), \nonumber \\
&\lambda_4=\left(%
\begin{array}{ccc}
  0 & 0 & 1 \\
  0 & 0 & 0 \\
  1 & 0 & 0 \\
\end{array}%
\right), \nonumber
    &\lambda_5=\left(%
\begin{array}{ccc}
  0 & 0 & -i \\
  0 & 0 & 0 \\
  i & 0 & 0 \\
\end{array}%
\right), \quad \quad
&\lambda_6=\left(%
\begin{array}{ccc}
  0 & 0 & 0 \\
  0 & 0 & 1 \\
  0 & 1 & 0 \\
\end{array}%
\right), \\
&\lambda_7=\left(%
\begin{array}{ccc}
  0 & 0 & 0 \\
  0 & 0 & -i \\
  0 & i & 0 \\
\end{array}%
\right), \quad
&\lambda_8=\frac{1}{\sqrt{3}}\left(%
\begin{array}{ccc}
  1 & 0 & 0 \\
  0 & 1 & 0 \\
  0 & 0 & -2 \\
\end{array}%
\right)
\nonumber\\
\end{align}
and $\n$ is now an eight component coherence vector. Note that our
representation of the state in terms of Pauli and Gell-Mann
matrices is not unique. Any other appropriate basis will be
related to this basis through an orthogonal matrix transformation
that would be $3$ and $8$ dimensional respectively \cite{kimura}.
For arbitrary dimensions, the density matrix is given by:
\begin{equation}\label{}
  \rho=\frac{1}{d}(\mathbb{I}+\sqrt\frac{d(d-1)}{2}\n\cdot\lambda)
\end{equation}
where $\n$ is the $d^2-1$ element coherence vector and $\lambda$
are $d\times d$ matrices satisfying the Lie algebra of SU(d)
\cite{byrd}. The mixedness is introduced in the coherence vectors
by $\n=p_1\n_1+p_2\n_2+p_3\n_3$ where $\n_i$ are the coherence
vectors corresponding to the $i$th state.

\begin{figure}[h]
\begin{center}
\rotatebox{0}{\resizebox{!}{6cm}{\includegraphics{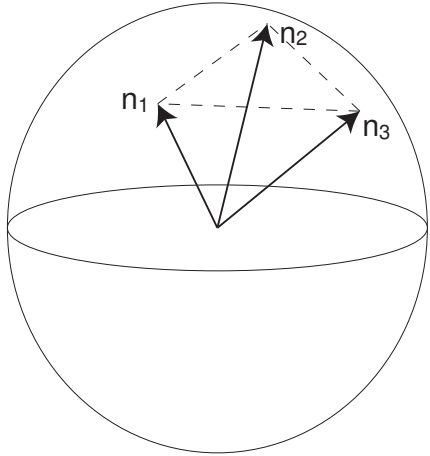}}}
\caption{\label{blochjul3} Coherence vector space with three
general states. The dotted lines denote the perimeter. Note that
in two dimensions, the ball itself is a Bloch sphere and every
point corresponds to a physical state. But for higher dimensions,
the coherence vector space is a proper subset of the ball.
\cite{kimura}.}
\end{center}
\end{figure}

\begin{figure}[h]
\begin{center}
\rotatebox{0}{\resizebox{!}{6cm}{\includegraphics{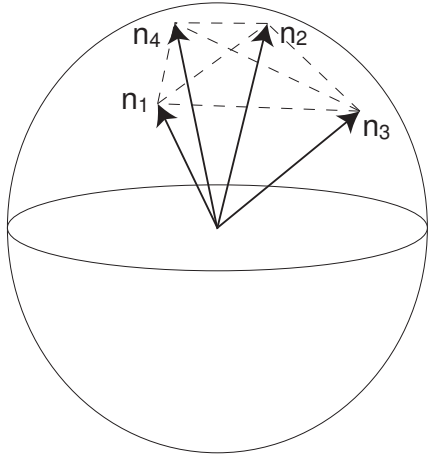}}}
\caption{\label{blochjul4}Coherence vector space with four states.
The dotted lines again denote the perimeter.}
\end{center}
\end{figure}

We now introduce a quantity called the perimeter, $P$, defined as:
\begin{eqnarray}\label{perimeter}
    P&=&|\n_1-\n_2|^2+|\n_2-\n_3|^2+|\n_3-\n_1|^2 \\
    &=&
    2\n_1^2+2\n_2^2+2\n_3^2-2\n_1\cdot\n_2-2\n_2\cdot\n_3-2\n_3\cdot\n_1
    \\
    &=& 6-2(\n_1\cdot\n_2+\n_2\cdot\n_3+\n_3\cdot\n_1)
\end{eqnarray}
This quantity tells us how different the three states are on
average. The larger the perimeter, the more orthogonal the states
become. Note that this quantity is related to the sum of the
overlaps of the quantum states (for example in three dimensions):
\begin{eqnarray}\label{Q}
  Q &=& |\langle\psi_1|\psi_2\rangle|^2+|\langle\psi_2|\psi_3\rangle|^2+|\langle\psi_3|\psi_1\rangle|^2 \\
   &=& Tr(\rho_1\rho_2)+Tr(\rho_2\rho_3)+Tr(\rho_3\rho_1) \\
   &=& \frac{1}{3}(1+2\n_1\cdot\n_2)+\frac{1}{3}(1+2\n_2\cdot\n_3)+\frac{1}{3}(1+2\n_3\cdot\n_1) \\
   &=& 1+\frac{6-P}{3}
\end{eqnarray}
The negative sign makes sense because the more/less parallel the
states are, the smaller/larger the perimeter. If the states are
identical, $P=0$ (since $\n_i\cdot\n_i=1)$ and if they are
orthogonal, $P=9$ (since $\n_i\cdot\n_j=-1/(d-1)$ for $i\neq j$
where $d$ is the dimension of the system). Note that this is for
the three dimensional case. With three states, we can visualize
the perimeter as the square distances of each side of a triangle
with each vertex representing a quantum state (see Figure
\ref{blochjul3}).
As soon as we consider more states, the usual meaning of perimeter
breaks down because we must include more than two distances for
each state. For example with four states, we will have the square
distances of each side of a four sided polygon as well as the two
lines adjoining opposite vertices (see Figure \ref{blochjul4}).
Hence the term "average distinguishability" may be more
appropriate than perimeter but we will continue to use the latter
throughout the paper. We would now expect, as mentioned earlier in
the introduction, that the larger the perimeter, the more
distinguishable (orthogonal) the states comprising the mixture,
and the higher the value of the entropy. This is, as will be shown
in more detail soon, true for qubits, but fails in higher
dimensions in general.

\begin{figure}[h]
\begin{center}
\rotatebox{0}{\resizebox{!}{6cm}{\includegraphics{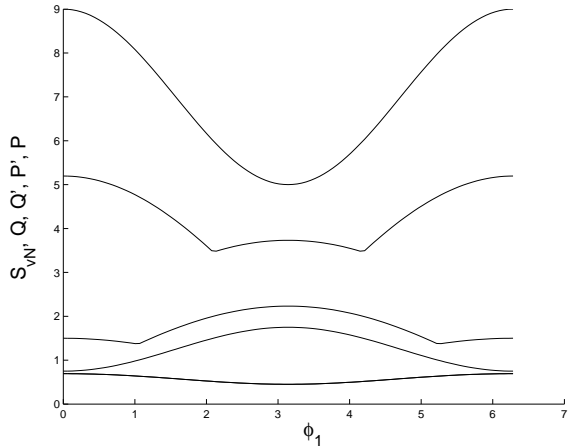}}}
\caption{\label{pqanomaly} The bottom line shows the von Neumann
entropy, the one above shows $Q$ and so on until the top line
shows $P$. $P'$ and $Q'$ are without the squares. Note that just
less than $\phi_1=\pi$, $S_{vN}$ decreases as $P'$ increases. Also
around $\phi_1=0$ and $\phi_1=2\pi$, $S_{vN}$ decreases as $Q'$
decreases and $S_{vN}$ increases as $Q'$ increases respectively.}
\end{center}
\end{figure}

Before we go into the main topic of the paper, we point out
another important issue. Namely, changing the definition of the
perimeter and $Q$ by removing the squares in equations
(\ref{perimeter}) and (\ref{Q}) respectively, changes the
behaviour of the perimeter with respect to the entropy. In
particular, we can now observe an increase in entropy by
decreasing the perimeter (or equivalently increasing the overlap)
for the two dimensional ensemble contrary to \cite{jozsaschlienz}.
Let us consider the following states:
\begin{eqnarray}
  |\psi_1\rangle &=& \cos(\theta_1/2)|0\rangle+\exp(-i\phi_1)\sin(\theta_1/2)|1\rangle \\
  |\psi_2\rangle &=& \cos(\theta_2/2)|0\rangle+\exp(-i\phi_2)\sin(\theta_2/2)|1\rangle \\
  |\psi_3\rangle &=& \cos(\theta_3/2)|0\rangle+
  \exp(-i\phi_3)\sin(\theta_3/2)|1\rangle
\end{eqnarray}
or equivalently the following coherence vectors:
\begin{eqnarray}
  \n_1 &=& [\sin(\theta_1)\cos(\phi_1),\sin(\theta_1)\sin(\phi_1),\cos(\theta_1)] \\
  \n_2 &=& [\sin(\theta_2)\cos(\phi_2),\sin(\theta_2)\sin(\phi_2),\cos(\theta_2)] \\
  \n_3 &=& [\sin(\theta_3)\cos(\phi_3),\sin(\theta_3)\sin(\phi_3),\cos(\theta_3)]
\end{eqnarray}
Let us fix $\theta_i=\pi/2$, $\phi_2=2\pi/3$, $\phi_3=4\pi/3$ and
vary $\phi_1$ from $0\to 2\pi$. In the Bloch sphere picture, the
states lie on the equator with each state initially equally
spaced. $\psi_1$ or $\n_1$ rotates around once remaining on the
equatorial plane while keeping the other two states fixed. Figure
\ref{pqanomaly} shows the anomaly.
Since this behaviour is counterintuitive, we will hereafter
continue to use the original definitions of $P$ and $Q$ because
they avoid the above anomaly and allow a simple relationship
between the perimeter and the total overlap. So, in summary, we
now have that the larger the perimeter (or equivalently the
smaller the $Q$), the larger the von Neumann entropy keeping all
other variables constant.

The geometric phase is a phase that is observed when a state
evolves in parameter space (e.g. the parameter could be a magnetic
field strength) \cite{berry}. A more amenable interpretation for
our present purposes is the quantum version of the Pancharatnam
relative phase \cite{pancharatnam}. See \cite{sjokvist} for a
concise modern introduction. We can calculate the geometric phase
$\gamma$ by:
\begin{equation}\label{geometricphase}
    \gamma_{ijk}=\arg\{Tr(|\psi_i\rangle\langle\psi_i|\psi_j\rangle\langle\psi_j|\psi_k\rangle\langle\psi_k|)\}
\end{equation}
For three states in two dimensions, we get:
\begin{equation}\label{geometricphase2}
    \tan\gamma_{123}=\frac{\n_1\times\n_2\cdot\n_3}{1+\n_1\cdot\n_2+\n_2\cdot\n_3+\n_3\cdot\n_1}
\end{equation}
where $\n_1\times\n_2$ is the ordinary cross product. For three
states in three dimensions, we get \cite{arvind1}:
\begin{equation}\label{geometricphase3}
    \tan\gamma_{123}=\frac{2\sqrt{3}\n_1\cdot\n_2\wedge\n_3}{(\n_1+\n_2+\n_3)^2+2\n_1\cdot\n_2\star\n_3-2}
\end{equation}
where $\n_{1}\cdot\n_{2}\wedge\n_{3}=n_{1i}f_{ijk}n_{2j}n_{3k}$
and $\n_1\cdot\n_2\star\n_3=\sqrt{3}n_{1i}d_{ijk}n_{2j}n_{3k}$.
$i, j, k$ refer to the components of the vectors, $f_{ijk}$ are
the antisymmetric $SU(3)$ structure constants and $d_{ijk}$ are
the symmetric tensors. Note that $\gamma_{123}$ refers to the
phase taking three states of any dimensionality. Exact definitions
and other useful formulae are given in \cite{arvind1} but for
convenience, we state them here:
\begin{equation}\label{}
    [\lambda_i,\lambda_j]=2if_{ijk}\lambda_k
\end{equation}
\begin{equation}\label{}
    \{\lambda_i,\lambda_j\}=\frac{4}{3}\delta_{ij}+2d_{ijk}\lambda_k
\end{equation}
\begin{equation}\label{}
    f_{123}=1,f_{458}=f_{678}=\frac{\sqrt{3}}{2},f_{147}=f_{246}=f_{257}=f_{345}=f_{516}=f_{637}=\frac{1}{2}
\end{equation}
\begin{equation}\label{}
    d_{118}=d_{228}=d_{338}=-d_{888}=\frac{1}{\sqrt{3}},d_{448}=d_{558}=d_{668}=d_{778}=-\frac{1}{2\sqrt{3}}
    \nonumber
\end{equation}
\begin{equation}\label{}
    d_{146}=d_{157}=-d_{247}=d_{256}=d_{344}=d_{355}=-d_{366}=-d_{377}=\frac{1}{2}
\end{equation}
Other useful formulae are:
\begin{equation}\label{}
    \lambda_i\lambda_j=\frac{2}{3}\delta_{ij}+(d_{ijk}+if_{ijk})\lambda_k
\end{equation}
\begin{equation}\label{}
    Tr\lambda_i=0, Tr(\lambda_i\lambda_j)=2\delta_{ij}
\end{equation}

The visibility is defined by:
\begin{equation}\label{visibility}
    V_{ijk}=|Tr(|\psi_i\rangle\langle\psi_i|\psi_j\rangle\langle\psi_j|\psi_k\rangle\langle\psi_k|)|
\end{equation}
where the name originates from $V$ corresponding to how visible or
how large the amplitude is in an interferometer
\cite{ekert,sjokvist2}. Note that $V\cos\gamma$ is equal to the
denominator of $\tan\gamma$ given in equations
(\ref{geometricphase2}) and (\ref{geometricphase3}) for two and
three dimensions respectively.

\section{Results}
In this section we obtain the following results. We first show
that in two dimensions, the entropy depends on either the
perimeter or the product of the visibility and the cosine of the
geometric phase but not both together. We next show that for three
states in three dimensions, we need both quantities and nothing
else. The same is shown to be true for three states in any
dimension as expected. Then we show that for many states in three
dimensions, the entropy depends now on the perimeter and all the
possible combinations of the product of the visibility and the
cosine of the geometric phase for every triplet of states. In the
last subsection, we generalize to any dimensions and any number of
states by using a closed-form solution of the entropy obtained by
Chumakov et. al. \cite{chumakov}. Note that our results will apply
for general mixtures of pure states, i.e. unequal probabilities,
however, we will frequently express them with equal probabilities
for convenience.

\subsection{Any Number of States in Two Dimensions} 
\begin{figure}[h]
\begin{center}
\rotatebox{0}{\resizebox{!}{6cm}{\includegraphics{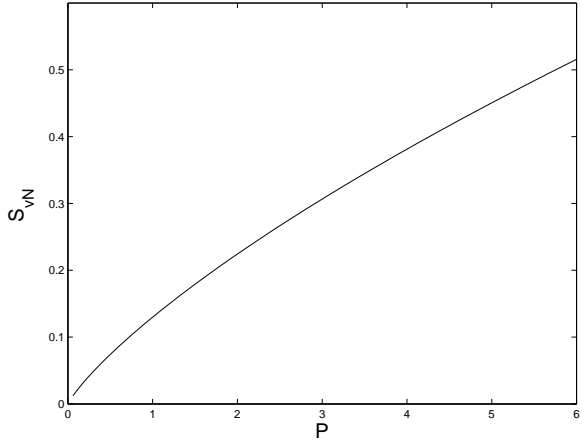}}}
\caption{\label{vne2vp} von Neumann entropy versus perimeter in
two dimensions for $t=3$}
\end{center}
\end{figure}

As is shown in
\cite{jozsaschlienz}, entropy cannot be increased by increasing
the average overlap of the ensemble in two dimensions for any
number of states. We will show this by giving an explicit formula
of von Neumann entropy as a function of perimeter. We can also
rewrite it as a function of geometric phase (modulated by the
visibility as will be defined later in this section) but the three
quantities will not appear together in the function. We must first
find the eigenvalues $x_{\pm}$ of the density operator which will
give $S_{vN}=-x_+\ln x_+-x_-\ln x_-$. Introduce
$\n=\frac{1}{t}(\n_1+\n_2+\ldots+\n_t)=(n_1,n_2,n_3)$ where
$n_i=\frac{1}{t}(n_{1i}+n_{2i}+n_{3i}+\ldots+n_{ti})$. The first
subscript refers to the state, the second subscript refers to the
vector component and $t$ denotes the number of states. The
eigenvalues are $x_{\pm}=(1\pm\sqrt{n_1^2+n_2^2+n_3^2})/2$. Note
that $n_1^2+n_2^2+n_3^2=\n\cdot\n$. Since
$\n\cdot\n=(t+2\n_1\cdot\n_2+2\n_2\cdot\n_3+2\n_3\cdot\n_1+\ldots+2\n_{t-1}\cdot\n_t)/t^2$
and generalizing the definition of $P$ above to $t$ states ($6$
becomes $t(t-1)$), $\n\cdot\n=(t^2-P)/t^2$. This gives:
\begin{equation}\label{vne2}
    S_{vN}=-\frac{1+\sqrt{\frac{t^2-P}{t^2}}}{2}\ln\frac{1+\sqrt{\frac{t^2-P}{t^2}}}{2}-\frac{1-\sqrt{\frac{t^2-P}{t^2}}}{2}\ln\frac{1-\sqrt{\frac{t^2-P}{t^2}}}{2}
\end{equation}
Figure \ref{vne2vp} plots this.
We see that the von Neumann entropy is a monotonically increasing
function of perimeter. Using equation (\ref{geometricphase2}), we
can also write $P=8-2V_{123}\cos\gamma_{123}$ for $t=3$ where
$V_{123}\cos\gamma_{123}=1+\n_1\cdot\n_2+\n_2\cdot\n_3+\n_3\cdot\n_1$.
We find that in two dimensions, increasing the geometric phase
corresponds to an increase in entropy (negative values of
$\cos\gamma$ become unphysical since at most $P=9$ and this
corresponds to all three states being orthogonal which is not
possible in two dimensions). Likewise, decreasing $V$ corresponds
to an increase in $P$ and hence entropy. For larger number of
states, we can define the perimeter as a function of
$V_{ijk}\cos\gamma_{ijk}$ for all the combinations of three
states:
\begin{equation}\label{}
    P=t(t-1)+\frac{2(t!)}{3!(t-3)!(t-2)}-\frac{2\Gamma}{t-2}
\end{equation}
where
\begin{equation}\label{}
    \Gamma=\sum_{k>j>i=1}^tV_{ijk}\cos\gamma_{ijk}
\end{equation}
We can see that even for larger number of states in two
dimensions, since the perimeter increases as geometric phase
increases, the von Neumann entropy also increases. Note that in
the case of unequal prior probabilities, there is no
straightforward method of relating the perimeter to the geometric
phase unless we redefine the geometric phase by incorporating the
unequal prior probabilities.

The above calculations were for equal prior probabilities
($p=1/t$). If we consider instead unequal prior probabilities
$p_i$, we must redefine the perimeter:
\begin{equation}\label{}
    \tilde{P}=|p_1\n_1-p_2\n_2|^2+|p_2\n_2-p_3\n_3|^2+|p_3\n_3-p_1\n_1|^2+\ldots+|p_t\n_t-p_{t-1}\n_{t-1}|^2
    \nonumber
\end{equation}
and then we obtain:
\begin{equation}\label{}
    \n\cdot\n=t\sum_{j=1}^tp_j^2-\tilde{P}
\end{equation}
and therefore entropy is still just a function of the perimeter.
We can see that this reduces to the equal prior probability result
above and when $p_i=1$, $\n\cdot\n=1$ thus $S_{vN}=0$, as it
should be for a pure state.

\subsection{Three States in Three Dimensions}
\begin{figure}[h]
\begin{center}
\rotatebox{0}{\resizebox{!}{6cm}{\includegraphics{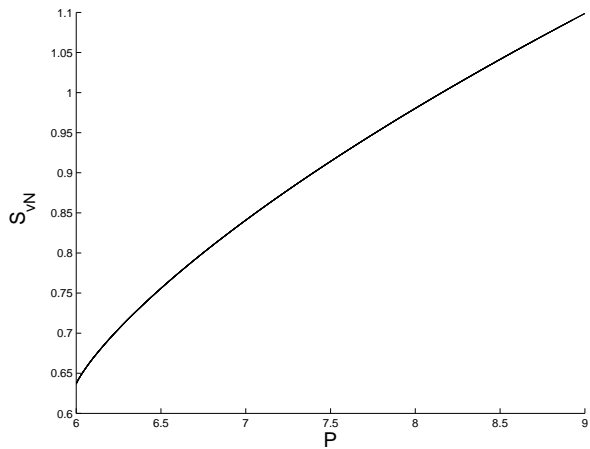}}}
\caption{\label{vne3vpconstVgamma} von Neumann entropy versus
perimeter in three dimensions}
\end{center}
\end{figure}

Similar steps are taken as the previous subsection. First we
introduce the coherence vector
$\n=(n_1,n_2,n_3,n_4,n_5,n_6,n_7,n_8)=\frac{1}{3}(\n_1+\n_2+\n_3)$
with equal prior probabilities. The perimeter is given by
$\n\cdot\n=\frac{1}{9}(9-P)$. Our density matrix is now a three by
three matrix, and in order to compute its entropy we need to be
able to find the eigenvalues first. This leads to solving the
following cubic equation:
\begin{equation}\label{cubicequation}
    x^3+Ax^2+Bx+C=0
\end{equation}
where
\begin{eqnarray}
  A &=& -1 \\
  B &=& \frac{1-\n\cdot\n}{3} \\
  C &=& \frac{\n\cdot\n}{9}-\frac{1}{27}-\frac{2}{
  27}V_{123}\cos\gamma_{123}
\end{eqnarray}
with
$V_{123}\cos\gamma_{123}=(\n_1+\n_2+\n_3)^2+2\n_1\cdot\n_2\star\n_3-2$.
The solution is given by \cite{mathworld,nickalls}:
\begin{eqnarray}
  x_1 &=& 2\sqrt{-T}\cos\frac{\theta}{3}-\frac{A}{3} \nonumber\\
  x_2 &=& 2\sqrt{-T}\cos\frac{\theta+2\pi}{3}-\frac{A}{3} \nonumber\\
  x_3 &=& 2\sqrt{-T}\cos\frac{\theta+4\pi}{3}-\frac{A}{3} \label{x1x2x3}
\end{eqnarray}
where
\begin{eqnarray}
  R &=& \frac{9AB-27C-2A^3}{54}=\frac{V_{123}\cos\gamma_{123}}{27} \\
  T &=& \frac{3B-A^2}{9}=-\frac{\n\cdot\n}{9} \\
  \theta &=& \arccos\frac{R}{\sqrt{-T^3}}=\arccos\frac{V_{123}\cos\gamma_{123}}{(\n\cdot\n)^{3/2}}
\end{eqnarray}
The von Neumann entropy is:
\begin{equation}
    S_{vN}=-x_1\ln x_1-x_2\ln x_2-x_3\ln x_3 \label{vne3}
\end{equation}
Let us look at a couple of examples. The first example uses the
three states given in \cite{arvind1}:
\begin{eqnarray}
  |\psi_1\rangle &=& |2\rangle \\
  |\psi_2\rangle &=& \sin\xi|1\rangle+\cos\xi|2\rangle \\
  |\psi_3\rangle &=&
  \sin\eta\cos\zeta|0\rangle+e^{i\chi}\sin\eta\sin\zeta|1\rangle+\cos\eta|2\rangle
\end{eqnarray}
where $0\leq\xi,\eta,\zeta\leq\pi/2$ and $0\leq\chi<2\pi$. By
setting $\xi=\pi/2$, $\chi=0$ and $\eta=\pi/2$, we can set the
geometric phase $\gamma$ and visibility $V$ to vanish. Then by
varying $\zeta$, we can observe the dependence of von Neumann
entropy $S_{vN}$ on perimeter $P$ between $6\leq P\leq 9$ as is
shown in Figure \ref{vne3vpconstVgamma}.
As is the case in two dimensions (Figure \ref{vne2vp}), $S_{vN}$
increases monotonically when $P$ increases. Note that $S_{vN}$ is
bounded by the maximum entropy allowable in a $d$ dimensional
system $S_{max}=\ln d$. In fact the monotonically increasing
property can be explicitly checked by differentiating equation
(\ref{vne3}) with respect to $P$ and realizing the result to be
positive for the above range of $P$. In order to inspect smaller
values of $P$, we must also vary $V$ or $\gamma$.

\begin{figure}[h]
\begin{center}
\rotatebox{0}{\resizebox{!}{6cm}{\includegraphics{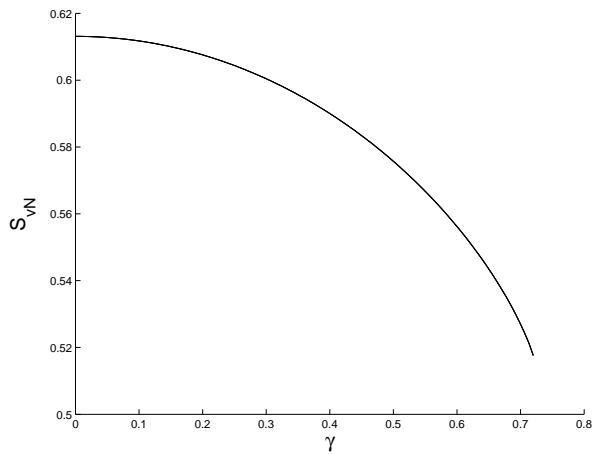}}}
\caption{\label{vne3vgammaconstVp} von Neumann entropy versus
geometric phase in three dimensions}
\end{center}
\end{figure}

The second example uses the three states given in
\cite{jozsaschlienz}:
\begin{eqnarray}
  |\psi_1\rangle &=& |0\rangle \\
  |\psi_2\rangle &=& \frac{1}{\sqrt{2}}|0\rangle+\frac{1}{\sqrt{2}}|1\rangle \\
  |\psi_3\rangle &=&
  \frac{1}{\sqrt{3}}|0\rangle+\frac{2e^{i\gamma}-1}{\sqrt{3}}|1\rangle+\sqrt{\frac{4}{3}\cos\gamma-1}|2\rangle
\end{eqnarray}
where $\gamma$ is the geometric phase which is bounded here by
$\gamma_{max}=0.72$ radians. These three states keep $V$ and $P$
fixed so that we can inspect how $S_{vN}$ depends on $\gamma$
alone. By calculating the perimeter, geometric phase and
visibility and using equation (\ref{vne3}), we obtain Figure
\ref{vne3vgammaconstVp} which is identical to the graph given in
\cite{jozsaschlienz}.
In contrast to the two dimensional case where we can only
increase/decrease the entropy as we increase/decrease the
geometric phase, we also have that the entropy decreases/increases
as the geometric phase increases/decreases.

Why should there be a transition between the two to the three
dimensional case? Mathematically speaking, the distinctions are
that the symmetric tensor does not exist in two dimensions and the
isomorphism between $SU(2)$ and $SO(3)$ does not exist between
$SU(3)$ and $SO(8)$. This means that in the eight dimensional
ball, there are patches corresponding to unphysical states and
therefore our intuition of what the phase is geometrically as well
as how the perimeter changes is lost. A precise formulation of
this remark is left for future research.

\subsection{Three States in Any Dimensions}
We can increase the number of dimensions arbitrarily by
considering three general states $|\alpha\rangle$, $|\beta\rangle$
and $|\delta\rangle$. We can construct the density operator with
equal prior probabilities,
$\rho=\frac{1}{3}(|\alpha\rangle\langle\alpha|+|\beta\rangle\langle\beta|+|\delta\rangle\langle\delta|)$.
As the above states are not orthogonal, we use the Gram-Schmidt
procedure to obtain the following orthogonal states:
\begin{eqnarray}
  |v_1\rangle &=& \frac{|\alpha\rangle}{\||\alpha\rangle\|}=|\alpha\rangle \\
  |v_2\rangle &=& \frac{|\beta\rangle-\langle v_1|\beta\rangle|v_1\rangle}{\||\beta\rangle-\langle v_1|\beta\rangle|v_1\rangle\|}=\frac{|\beta\rangle-\langle v_1|\beta\rangle|v_1\rangle}{\sqrt{1-|\langle\alpha|\beta\rangle|^2}} \\
  |v_3\rangle &=& \frac{|\delta\rangle-\langle v_2|\delta\rangle|v_2\rangle-\langle v_1|\delta\rangle|v_1\rangle}{\||\delta\rangle-\langle v_2|\delta\rangle|v_2\rangle-\langle
  v_1|\delta\rangle|v_1\rangle\|}=\frac{|\delta\rangle-\langle v_2|\delta\rangle|v_2\rangle-\langle
  v_1|\delta\rangle|v_1\rangle}{\sqrt{1-|\langle v_2|\delta\rangle|^2-|\langle\alpha|\delta\rangle|^2}}
\end{eqnarray}
We can invert these and substitute into the density operator. By
noting that
$Q=|\langle\alpha|\beta\rangle|^2+|\langle\beta|\delta\rangle|^2+|\langle\delta|\alpha\rangle|^2$
(used instead of perimeter) and
$V\cos\gamma=\Re\{\langle\alpha|\delta\rangle\langle\delta|\beta\rangle\langle\beta|\alpha\rangle\}$
we find that:
\begin{eqnarray}
  A &=& -1 \\
  B &=& \frac{3-Q}{9} \\
  C &=& \frac{-1+Q-2V_{123}\cos\gamma_{123}}{27}
\end{eqnarray}
which give:
\begin{eqnarray}
  R &=& \frac{V_{123}\cos\gamma_{123}}{27} \\
  T &=& -\frac{Q}{27} \\
  \theta &=&
  \arccos\frac{\sqrt{27}V_{123}\cos\gamma_{123}}{Q^{3/2}}
\end{eqnarray}
By noting that $\n\cdot\n=1-P/9$ and $Q=(9-P)/3$ gives
$\n\cdot\n=Q/3$, we find that these equations are identical to the
equivalent ones appearing in the previous subsection. This is not
surprising because three states in arbitrary dimensions can be
represented by a rank three matrix. The important point to observe
is that we do not require any other quantity to define entropy. We
still only require the total overlap (or perimeter), geometric
phase and visibility.

For unequal prior probabilities, we do not get a simple
generalization as in the two dimensional case because we have a
mixture of probabilities to the power of two and to the power of
three which cannot be factored out. We have explicitly:
\begin{eqnarray}
  A &=& -1 \\
  B &=& p_1p_2+p_2p_3+p_1p_3-p_1p_2|\langle\alpha|\beta\rangle|^2-p_2p_3|\langle\beta|\delta\rangle|^2-p_1p_3|\langle\alpha|\delta\rangle|^2 \\
  C &=& p_1p_2p_3(-1+Q-2V_{123}\cos\gamma_{123})
\end{eqnarray}
They reduce to the equal prior probability case and for a pure
state, $B=C=0$ therefore when substituting into equation
(\ref{x1x2x3}), we obtain the desired $S_{vN}=0$. It is
interesting to note that now $R$ contains the overlap as well as
the visibility and geometric phase, hence altering the above form
of entropy. Since the probabilities directly influence how mixed
the ensemble is, it is not surprising that the form of the entropy
should change.

\subsection{Any Number of States in Three Dimensions} We have so
far looked at only three states in effectively three dimensions.
We will now consider the three dimensional case with $N$ number of
states. We now have $\n=(n_1,n_2,\ldots
,n_8)=\frac{1}{N}(\n_1+\n_2+\ldots+\n_{N-1}+\n_N)$. The perimeter
can be written in a compact form as:
\begin{equation}\label{}
  P=N(N-1)-2\sum_{i>j=1}^N\n_i\cdot\n_j
\end{equation}
Note that this is also true for any dimensions. Another useful
formula is:
\begin{equation}\label{}
  \n\cdot\n=\frac{1}{N^2}[N+2\sum_{i>j=1}^N\n_i\cdot\n_j]
\end{equation}
also true for any dimensions. We know that the cubic coefficients
$A$ and $B$ remain the same as before. $C$ is the only one that
needs to be modified. We find that:
\begin{multline}\label{}
  C=\frac{\n\cdot\n}{9}-\frac{1}{27}-\frac{2}{9N^3}\sum_{i>j>k=1}^NV_{ijk}\cos\gamma_{ijk}
  \\
  +\frac{2}{9N^3}\Bigg(\frac{N!}{(N-3)!3!}+(N-3)\Big(N(N-1)-P\Big)-\frac{N}{3}\Bigg)
\end{multline}
Notice that when $N=3$, this reduces to the aforementioned three
state case. It is interesting to note that the geometric phase
term still contains only three states albeit with all the possible
combinations of three states.

\subsection{Any Number of States in any Dimension}
We know that the von Neumann entropy can be expanded in a power
series $S_{vN}=\sum_{i=1}^\infty c_iTr\rho^i$ where $c_i$ are the
expansion coefficients. The exact values of the $c_i$'s are not
relevant for our discussion. Keyl and Werner \cite{keylwerner}
have shown that in order to calculate the eigenvalues of a $d$
dimensional density matrix, it is necessary and sufficient to
obtain all the traces of the powers of the density matrix up to
the $d$th power. $Tr\rho^2$ contains the perimeter and $Tr\rho^3$
contains the geometric phase with three states as is shown in the
next section. With a $d$ dimensional system, the entropy will
contain geometric phase terms up to $d$ states. Obtaining a
closed-form solution of the entropy for higher than four
dimensions is difficult because there is no equation using only
radicals to solve the quintic or higher equation. However,
Chumakov et. al. \cite{chumakov} have a closed-form solution for
arbitrary dimensional systems which requires traces of powers of
the density matrices up to $d$. In turn, these traces contain only
the perimeter and all combinations of the product of visibility
and the cosine of the geometric phase up to $d$ states, e.g.
$V_{ab\ldots d}\cos\gamma_{ab\ldots d}$. So we conclude that even
for higher dimensional systems, the entropy can be expressed as a
function of perimeter and the product of visibility and the cosine
of the geometric phases. Also, this is a natural way to view the
fact that entropy should be a function of perimeter and geometric
phases as will be also clear from the next section.

\section{Experimental Measurements of Entropy, \\ Perimeter and Phase}
We use the simple quantum network based on the controlled-SWAP
gate presented in \cite{ekert} which extracts properties of
quantum states bypassing the need for quantum tomography.
Physically, the network is a representation of the Mach-Zehnder
interferometer \cite{vedral}.

Since we have shown the von Neumann entropy as a function of
perimeter (overlap), geometric phase and visibility, we can
experimentally measure this entropy by calculating $Tr(\rho^2)$
for the perimeter and $Tr(\rho^3)$ for the visibility and
geometric phase where $\rho=\frac{1}{3}(\rho_1+\rho_2+\rho_3)$
with equal prior probabilities for three states in three
dimensions. However, we can generalize this experimental procedure
for any dimensions and any number of states by calculating the
traces of up to the $d$th power of the density matrix
$\rho=\frac{1}{N}\sum_{i=1}^N\rho_i$ where $N$ is the number of
states. Consider a setup with two separable subsystems
$\rho\otimes\rho$ and three separable subsystems
$\rho\otimes\rho\otimes\rho$. We now introduce the swap operator
$W$, $W|a\rangle\otimes|b\rangle=|b\rangle\otimes|a\rangle$ and
the shift operator $F$,
$F|a\rangle\otimes|b\rangle\otimes|c\rangle=|c\rangle\otimes|a\rangle\otimes|b\rangle$
for any pure states $|a\rangle$, $|b\rangle$ and $|c\rangle$. The
experimental procedure which will be described shortly measures
$TrW(\rho\otimes\rho)=Tr(\rho^2)$ \cite{ekert} and similarly
$TrF(\rho\otimes\rho\otimes\rho)=Tr(\rho^3)$. This can be readily
generalized to the $d$th power of $\rho$ using the general shift
operator $S$ where
$S|a\rangle\otimes\ldots\otimes|c\rangle\otimes|d\rangle=|d\rangle\otimes|a\rangle\ldots\otimes|c\rangle$
so that $TrS(\rho^{\otimes d})=Tr(\rho^d)$.

\begin{figure}[h]
\begin{center}
\rotatebox{0}{\resizebox{!}{5cm}{\includegraphics{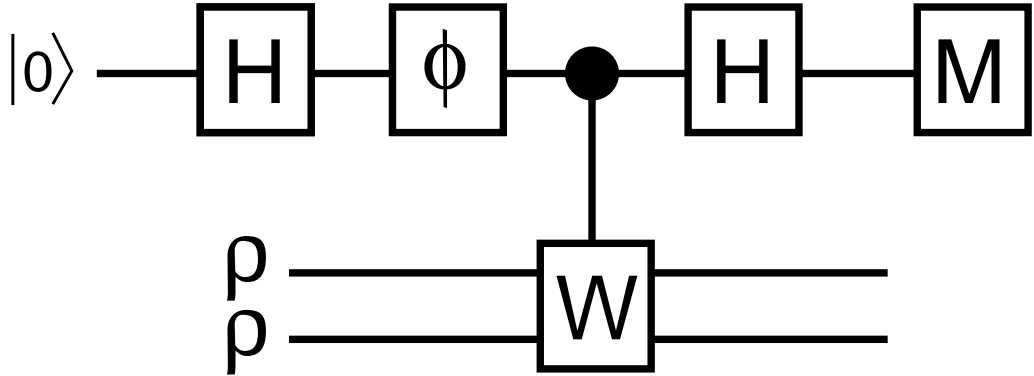}}}
\caption{\label{experiment} Experimental set up to ascertain
$Tr\rho^2$. We must exchange the swap operator $W$ with a shift
operator $S$ and input d $\rho$s to obtain $Tr\rho^d$. }
\end{center}
\end{figure}

We find on expansion:
\begin{eqnarray}
  Tr\rho^2 &=& \frac{1}{9}(3+2Tr\rho_1\rho_2+2Tr\rho_2\rho_3+2Tr\rho_1\rho_3)=\frac{1}{9}(3+2Q) \label{trrho2}\\
  Tr\rho^3 &=&
  \frac{1}{27}(3+6Tr\rho_1\rho_2+6Tr\rho_2\rho_3+6Tr\rho_1\rho_3+3Tr\rho_1\rho_2\rho_3+3Tr\rho_1\rho_3\rho_2)\nonumber\\
  &=& \frac{1}{27}(3+6Q+6V_{123}\cos\gamma_{123}) \label{trrho3}
\end{eqnarray}
The last line follows from
$Tr\rho_1\rho_2\rho_3=V_{123}e^{i\gamma_{123}}$ and
$Tr\rho_1\rho_3\rho_2=V_{123}e^{-i\gamma_{123}}$. Hence on
obtaining $Q$ and $V_{123}\cos\gamma_{123}$, we can calculate
$S_{vN}$ for three states in three dimensions. In principle, we
can also expand $Tr\rho^d$ to show that it contains $Q$ and all
the combinations of the product of visibility and the cosine of
the geometric phase. Figure \ref{experiment} shows the
experimental set up that may be used to measure the von Neumann
entropy (the diagram shows the case for two inputs of $\rho$ but
for a rank $d$ density matrix, we must inspect up to $d$ inputs of
$\rho$).
We will briefly describe how it calculates $Tr\rho^2$, and then
$Tr\rho^d$ is a straightforward extension. We begin with the
initial state $\rho_{in}=|0\rangle\langle
0|\otimes\rho\otimes\rho$. We apply the first Hadamard gate
$H=\frac{1}{2}\left(%
\begin{array}{cc}
  1 & 1 \\
  1 & -1 \\
\end{array}%
\right)$:
\begin{eqnarray}
  \rho_H &=& (H\otimes\mathbb{I}\otimes\mathbb{I})(|0\rangle\langle
0|\otimes\rho\otimes\rho)(H^\dag\otimes\mathbb{I}\otimes\mathbb{I}) \\
   &=& \frac{1}{2}\left(%
\begin{array}{cc}
  1 & 1 \\
  1 & 1 \\
\end{array}%
\right)\otimes\rho\otimes\rho
\end{eqnarray}
Then we apply the phase shift $\Phi=\left(%
\begin{array}{cc}
  e^{i\phi} & 0 \\
  0 & 1 \\
\end{array}%
\right)$ to get $\rho_\Phi=\frac{1}{2}\left(%
\begin{array}{cc}
  1 & e^{i\phi} \\
  e^{-i\phi} & 1 \\
\end{array}%
\right)\otimes\rho\otimes\rho$. Next is the controlled-swap
operation:
\begin{equation}\label{controlledswap}
    U_{cs}=\left(%
\begin{array}{cc}
  1 & 0 \\
  0 & 0 \\
\end{array}%
\right)\otimes\mathbb{I}\otimes\mathbb{I}+\left(%
\begin{array}{cc}
  0 & 0 \\
  0 & 1 \\
\end{array}%
\right)\otimes W
\end{equation}
and finally another Hadamard to obtain:
\begin{eqnarray}\label{}
    \rho_{out}&=&\frac{1}{4}\Bigg[\left(%
\begin{array}{cc}
  1 & 1 \\
  1 & 1 \\
\end{array}%
\right)\otimes\rho\otimes\rho+\left(%
\begin{array}{cc}
  1 & -1 \\
  -1 & 1 \\
\end{array}%
\right)\otimes W(\rho\otimes\rho)W^\dag \\
& & +e^{i\phi}\left(%
\begin{array}{cc}
  1 & -1 \\
  1 & -1 \\
\end{array}%
\right)\otimes(\rho\otimes\rho)W^\dag+e^{-i\phi}\left(%
\begin{array}{cc}
  1 & 1 \\
  -1 & -1 \\
\end{array}%
\right)\otimes W(\rho\otimes\rho)\Bigg] \nonumber
\end{eqnarray}
Since measuring the intensity $I$ is proportional to the
probability, we can measure in the computational basis $|0\rangle$
to get:
\begin{eqnarray}
 I &\propto& Tr\Big[|0\rangle\langle 0|\otimes\mathbb{I}\otimes\mathbb{I}\rho_{out}\Big] \nonumber\\
   &\propto& Tr\rho Tr\rho+Tr(W\rho\otimes\rho W^\dag)+e^{i\phi}Tr(\rho\otimes\rho W^\dag)+e^{-i\phi}Tr(W\rho\otimes\rho) \nonumber\\
   &=& 1+1+e^{i\phi}[TrW\rho\otimes\rho]^\ast+e^{-i\phi}TrW\rho\otimes\rho \\
   &=& 2+e^{i\phi}|Tr\rho^2|e^{-i\arg Tr\rho^2}+e^{-i\phi}|Tr\rho^2|e^{i\arg Tr\rho^2} \\
   &=& 2+2|Tr\rho^2|\cos[\phi-\arg Tr\rho^2]
\end{eqnarray}
We are able to adjust the phase $\phi$ so as to obtain the largest
intensity yielding $|Tr\rho^2|$ and $\phi=\arg Tr\rho^2$. Then we
acquire $Tr\rho^2=|Tr\rho^2|e^{i\arg Tr\rho^2}$. We also obtain
$Tr\rho^3$ following similar steps. We can obtain the von Neumann
entropy via (\ref{trrho2}) and (\ref{trrho3}) for three
dimensions. Naturally, we can calculate the von Neumann entropy
for $d$ dimensional systems by calculating the trace of the powers
of $\rho$ up to $d$ and utilizing the formula given in
\cite{chumakov}. So we see that the set up in Figure
\ref{experiment} allows us to measure both the entropy and the
product of the visibility and the cosine of the geometric phase.

\section{Summary and Conclusions}
We have explicitly shown the dependence of entropy on the
perimeter, geometric phase and the visibility. For an arbitrary
number of states in the two dimensional case, entropy is solely a
function of perimeter whereas for three states in three dimensions
and more states in higher dimensions, entropy is no longer just a
function of perimeter but also of geometric phase and visibility.
Finally we have shown a possible way to obtain the von Neumann
entropy experimentally. The same experimental interferometric set
up can also be used to measure the visibility and geometric phase
associated with a set of pure states. This clarifies why
physically the two seemingly unrelated concepts of entropy and
geometric phase should in fact depend on each other.

Finally, we would like to speculate on the possibility of the
geometric phase playing a role in black hole entropy
\cite{waldthermo}. It is well established that Black hole entropy
is proportional to the area of its event horizon. Similarly, in
the two dimensional Hilbert space, the geometric phase is given by
half the solid angle subtended by states involved. Moreover, we
have shown that the entropy in this case is only a function of the
geometric phase (modulated by the visibility), and as the phase
increases so does the entropy. So, we have the same kind of
behaviour as for black holes, namely that the larger the area the
larger the entropy. This dependence breaks down in higher
dimensions (as we have seen we can increase the phase and decrease
the entropy). Of course, the two areas do not live in the same
space. The black hole area arises from the physical boundary
separating the black hole from the rest of the universe whereas
the geometric phase is an area in Hilbert space. A very
interesting theme for future research would be to investigate if
this fact has any deeper significance and whether it implies that
the information in a black hole is made up of basic two
dimensional units (qubits) rather than higher dimensional units.
Alternatively we can explore the possibility of having states in
higher dimensional Hilbert spaces whose contribution to the
entropy from the perimeter would be small while dominated by the
geometric phase.

%

\textbf{Acknowledgements}. We would like to thank Arvind for
clarifying a few points about his paper,Angelo Carollo, Olaf
Dreyer, Damian Markham, Mio Murao, Soonmie Park and Andrei
Soklakov for fruitful discussions. We also acknowledge the
hospitality of Perimeter Institute where part of the work was
completed and the financial support from the Engineering and
Physical Sciences Research Council of UK, the European Union and
the Elsag S.p.a.

\bibliographystyle{unsrt}

\end{document}